\documentclass{bmcart}

\usepackage{amsthm,amsmath}
\usepackage[utf8]{inputenc} %unicode support
\usepackage{graphicx}
\usepackage[ruled,vlined]{algorithm2e}
\usepackage{natbib}
\usepackage{floatrow}
\usepackage[label font=bf,labelformat=simple]{subfig}
\usepackage{caption}
\usepackage{ulem} 

\startlocaldefs
\endlocaldefs

\begin{document}

%%% Start of article front matter
\begin{frontmatter}

\begin{fmbox}
\dochead{Research}


\title{A General Deep Learning Framework for Network Reconstruction and Dynamics Learning}

%%%%%%%%%%%%%%%%%%%%%%%%%%%%%%%%%%%%%%%%%%%%%%
%%                                          %%
%% Enter the authors here                   %%
%%                                          %%
%% Specify information, if available,       %%
%% in the form:                             %%
%%   <key>={<id1>,<id2>}                    %%
%%   <key>=                                 %%
%% Comment or delete the keys which are     %%
%% not used. Repeat \author command as much %%
%% as required.                             %%
%%                                          %%
%%%%%%%%%%%%%%%%%%%%%%%%%%%%%%%%%%%%%%%%%%%%%%

\author[
   addressref={aff1},                   % id's of addresses, e.g. {aff1,aff2}
%   corref={aff1},                       % id of corresponding address, if any
%   noteref={n1},                        % id's of article notes, if any
   email={3riccczz@gmail.com}   % email address
]{\inits{Z.Z.}\fnm{Zhang} \snm{Zhang}}
\author[
   addressref={aff3},
   email={yizhao.myriad@gmail.com} 
]{\inits{Y.Z.}\fnm{Yi} \snm{Zhao}}
\author[
   addressref={aff1},
   email={jing.liu@mail.bnu.edu.cn} 
]{\inits{J.L.}\fnm{Jing} \snm{Liu}}
\author[
   addressref={aff2},
   email={shawnwang.tech@gmail.com} 
]{\inits{S.W.}\fnm{Shuo} \snm{Wang}}
\author[
   addressref={aff4},
   email={try@feddy.org} 
]{\inits{S.W.}\fnm{Ruyi} \snm{Tao}}
\author[
   addressref={aff1},
   email={201621250011@mail.bnu.edu.cn} 
]{\inits{R-Y.X.}\fnm{Ruyue} \snm{Xin}}
\author[
   addressref={aff1},
   corref={aff1}, 
   email={zhangjiang@bnu.edu.cn}
]{\inits{J.Z.}\fnm{Jiang} \snm{Zhang}}

%%%%%%%%%%%%%%%%%%%%%%%%%%%%%%%%%%%%%%%%%%%%%%
%%                                          %%
%% Enter the authors' addresses here        %%
%%                                          %%
%% Repeat \address commands as much as      %%
%% required.                                %%
%%                                          %%
%%%%%%%%%%%%%%%%%%%%%%%%%%%%%%%%%%%%%%%%%%%%%%

% bnu
\address[id=aff1]{%                           % unique id
  \orgname{School of Systems Science, Beijing Normal University}, % university, etc
%   \street{Waterloo Road},                     %
  \postcode{100875}                                % post or zip code
  \city{Beijing},                             % city
  \cny{People’s Republic of China}                                    % country
}

% colorfulcloud
\address[id=aff2]{%                           % unique id
  \orgname{ColorfulClouds Pacific Technology Co., Ltd.}, % university, etc
  \street{No.04, Building C, 768 Creative Industrial Park, Compound 5A, Xueyuan Road, Haidian District},                     %
  \postcode{100083}                                % post or zip code
  \city{Beijing},                             % city
  \cny{People’s Republic of China}                                    % country
}

% nanjing university
\address[id=aff3]{%                           % unique id
  \orgname{School of Physics, Nanjing University}, % university, etc
%   \street{Waterloo Road},                     %
  \postcode{210093}                                % post or zip code
  \city{Nanjing},                             % city
  \cny{People’s Republic of China}          % country
}
% nanjing university
\address[id=aff4]{%                           % unique id
  \orgname{Swarma Campus (Beijing) Technology Co., Ltd}, % university, etc
%   \street{Waterloo Road},                     %
  \postcode{100083}                                % post or zip code
  \city{Beijing},                             % city
  \cny{People’s Republic of China}          % country
}

%%%%%%%%%%%%%%%%%%%%%%%%%%%%%%%%%%%%%%%%%%%%%%
%%                                          %%
%% Enter short notes here                   %%
%%                                          %%
%% Short notes will be after addresses      %%
%% on first page.                           %%
%%                                          %%
%%%%%%%%%%%%%%%%%%%%%%%%%%%%%%%%%%%%%%%%%%%%%%

% \begin{artnotes}
% %\note{Sample of title note}     % note to the article
% \note[id=n1]{Equal contributor} % note, connected to author
% \end{artnotes}

\end{fmbox}% comment this for two column layout

%%%%%%%%%%%%%%%%%%%%%%%%%%%%%%%%%%%%%%%%%%%%%%
%%                                          %%
%% The Abstract begins here                 %%
%%                                          %%
%% Please refer to the Instructions for     %%
%% authors on http://www.biomedcentral.com  %%
%% and include the section headings         %%
%% accordingly for your article type.       %%
%%                                          %%
%%%%%%%%%%%%%%%%%%%%%%%%%%%%%%%%%%%%%%%%%%%%%%

\begin{abstractbox}

\begin{abstract} % abstract
% \parttitle{First part title} %if any

Many complex processes can be viewed as dynamical systems on networks. However, in real cases, only the performances of the system are known, the network structure and the dynamical rules are not observed. Therefore, recovering latent network structure and dynamics from observed time series data are important tasks because it may help us to open the black box, and even to build up the model of a complex system automatically. Although this problem hosts a wealth of potential applications in biology, earth science, and epidemics etc., conventional methods have limitations. In this work, we introduce a new framework, Gumbel Graph Network (GGN), which is a model-free, data-driven deep learning framework to accomplish the reconstruction of both network connections and the dynamics on it. Our model consists of two jointly trained parts: a network generator that generating a discrete network with the Gumbel Softmax technique; and a dynamics learner that utilizing the generated network and one-step trajectory value to predict the states in future steps. We exhibit the universality of our framework on different kinds of time-series data: with the same structure, our model can be trained to accurately recover the network structure and predict future states on continuous, discrete, and binary dynamics, and outperforms competing network reconstruction methods.
%做了什么模型，达到了多少精确度
%动力学模型和网络重构

% \parttitle{Second part title} %if any
% Text for this section.
\end{abstract}

%%%%%%%%%%%%%%%%%%%%%%%%%%%%%%%%%%%%%%%%%%%%%%
%%                                          %%
%% The keywords begin here                  %%
%%                                          %%
%% Put each keyword in separate \kwd{}.     %%
%%                                          %%
%%%%%%%%%%%%%%%%%%%%%%%%%%%%%%%%%%%%%%%%%%%%%%


% MSC classifications codes, if any
%\begin{keyword}[class=AMS]
%\kwd[Primary ]{}
%\kwd{}
%\kwd[; secondary ]{}
%\end{keyword}

\end{abstractbox}
%
%\end{fmbox}% uncomment this for twcolumn layout

\end{frontmatter}

%%%%%%%%%%%%%%%%%%%%%%%%%%%%%%%%%%%%%%%%%%%%%%
%%                                          %%
%% The Main Body begins here                %%
%%                                          %%
%% Please refer to the instructions for     %%
%% authors on:                              %%
%% http://www.biomedcentral.com/info/authors%%
%% and include the section headings         %%
%% accordingly for your article type.       %%
%%                                          %%
%% See the Results and Discussion section   %%
%% for details on how to create sub-sections%%
%%                                          %%
%% use \cite{...} to cite references        %%
%%  \cite{koon} and                         %%
%%  \cite{oreg,khar,zvai,xjon,schn,pond}    %%
%%  \nocite{smith,marg,hunn,advi,koha,mouse}%%
%%                                          %%
%%%%%%%%%%%%%%%%%%%%%%%%%%%%%%%%%%%%%%%%%%%%%%

%%%%%%%%%%%%%%%%%%%%%%%%% start of article main body
% <put your article body there>

%%%%%%%%%%%%%%%%
%% Background %%
%%

\section{Introduction}
\label{section-intro}

Many complex processes can be viewed as dynamical systems on an underlying network structure. Network with the dynamics on it is a powerful approach for modeling a wide range of phenomena in real-world systems, where the elements are regarded as nodes and the interactions as edges\cite{albert2002statistical, strogatz2001exploring, newman2003structure}.  One particular interest in the field of network science is the interplay between the network topology and its dynamics\cite{boccaletti2006complex}. Much attention has been paid on how collective dynamics on networks are determined by the topology of graph. However, in real cases, only the performances, i.e., the time series of nodes states are observed, but the network structure and the dynamical rules are not known. Thus, the inverse problems, i.e., inferring network topology and dynamical rules based on the observed dynamics data, is more significant. This may pave a new way to detect the internal structure of a system according to its behaviors. Furthermore, it can help us to build up the dynamical model of a complex system according to the observed performance automatically. 

For example, inferring gene regulatory networks from expression data can help us to identify the major genes and reveal the functional properties of genetic networks\cite{gardner2003inferring}; in the study of climate changes, network reconstruction may help us to reveal the atmospheric teleconnection patterns and understand their underlying mechanisms\cite{boers2019complex}; it can also find applications in reconstructing epidemic spreading processes in social networks, which is essential to identifying the source and preventing further spreading\cite{shen2014reconstructing}. Furthermore, if not only the network structure but also the dynamics can be learned very well for these systems, surrogate models of the original problems can be obtained, on which, many experiments that are hard to implement on the original systems can be operated.  Another potential application is automated machine learning (AutoML)\cite{feurer2015efficient, quanming2018taking}. At present, the main research problem of Neural Architecture Search(NAS), a sub-area of AutoML, is to find the optimal neural network architecture in a space by the search strategy, and it is essentially a network reconstruction problem, in which the optimal neural network and the dynamical rules on it can be learned according to the observed training samples as time series.  In a word, reconstructions of network and dynamical rules are pivotal to a wide span of applications.

A considerable amount of methods have been proposed for reconstructing network from time series data. One class of them is based on the method of statistical inference such as Granger causality\cite{quinn2011estimating, brovelli2004beta}, and correlation measurements\cite{stuart2003gene, eguiluz2005scale, barzel2013network}. These methods, however, can usually discover functional connectivity and may fail to reveal structural connection \cite{feizi2013network}. This means that in the reconstructed system, strongly correlated areas in function need to be also directly connected in structure. Nevertheless this requirement is seldom satisfied in many real-world systems like brain \cite{park2013structural} and climate systems \cite{boers2019complex}. Another class of methods were developed for reconstructing structural connections directly under certain assumptions. For example, methods such as driving response\cite{timme2007revealing} or compressed sensing\cite{wang2011predicting, wang2011time, wang2011network, shen2014reconstructing} either require the functional form of the differential equations, or the target specific dynamics, or the sparsity of time series data. Although a model-free framework presented by Casadiego et al.\cite{casadiego2017model} do not have these limitations, it can only be applied to dynamical systems with continuous variables so that the derivatives can be calculated. Thus, a general framework for reconstructing network topology and learning dynamics from the time series data of various types of dynamics, including continuous, discrete and binary ones, is necessary.

Recently, deep Learning has gained success in many areas such as image classification \cite{krizhevsky2012imagenet} and speech recognition \cite{hinton2012deep}. Can we apply this state-of-the-art technique on network reconstruction problem? This is possible because Graph network framework \cite{battaglia2018relational} have enabled deep learning techniques applied on graph structures successfully by mapping graph-structured data onto Euclidean space with update and aggregation functions \cite{zonghanwu2019surveygnn}. With a wealth of different avenues available, GN can be tailored to perform various tasks, such as node or graph classification \cite{velivckovic2017graph,zhang2018gaan}, graph generation \cite{de2018molgan,li2018learning,bojchevski2018netgan,you2018graphrnn}, and spatial-temporal forecasting \cite{jain2016structural,li2017diffusion,yu2017spatio,yan2018spatial}. Recently, the topic of recovering interactions and predicting physical dynamics under given interaction networks has attracted much attention. A most used approach is introduced by Battaglia et al. \cite{battaglia2016interaction}, representing particles as nodes and interactions as edges, then reconstruct the trajectories in a inference process on the given graph. However, most of the works in this field have focused on physical reasoning task while few dedicate to solving the inverse problem of network science: revealing network topology from observed dynamics. Some related works \cite{vinmodel,permutationneuralnet} attempted to infer implicit interaction of the system to help with the state prediction via observation. But they did not specify the implicit interaction as the network topology of the system, therefore the network reconstruction task remains ignored. Of all literature as we known, only NRI (Neural Relational Inference) model\cite{kipf_neural_2018} is working on this goal. Nevertheless, only a few continuous dynamics such as spring model and Kuramoto model are studied, and discrete processes were never considered. So in the rest of this article, we will take NRI as one of our baselines and will be compared against our own model.

In this work we introduce Gumbel Graph Network (GGN), a model-free, data-driven method that can simultaneously reconstruct network topology and perform dynamics prediction from time series data of node states. It is able to attain high accuracy on both tasks under various dynamical systems, as well as multiple types of network topology. We first introduce our architecture which is called Gumbel Graph Networks in Section \ref{section-archi} and then give a brief overview of our experiments on three typical dynamics in Section \ref{section-exp}. In Section \ref{section-results}, we show our results. Finally, some concluding remarks and discussions are given in Section \ref{section-conclusion}.

\section{GGN Architecture}
\label{section-archi}

\subsection{Problem Overview}

The goal of our Gumbel Graph Network is to reconstruct the interaction graph and simulate the dynamics from the observational data of $N$ interacting objects.

Typically, we assume that the system dynamics that we are interested can be described by a differential equation $dX/dt = \psi(X^t, A)$ or the discrete iteration $ X^t=\psi(X^{t-1}, A)$, where ${ X }^{ t }=({ X }_{ 1 }^{ t },...,{ X }_{ N }^{ t })$ denotes the states of $N$ objects at time $t$, and ${ X }_{ i }$ is the state of the object $i$.  $\psi$ is the dynamical function, and $A$ is the adjacency matrix of an unweighted directed graph. However, $\psi$ and $A$ are unknown for us, and they will be inferred or reconstructed from a segment of time series data, i.e., $X=({ X }^{ t },...,{ X }^{ t+P })$, where $P$ is the number of prediction steps. 

Thus, our algorithm aims to learn the network structure (Specifically, the adjacency matrix) and the dynamical model $\psi$ simultaneously in an unsupervised way.

\subsection{Framework}

\begin{figure}[!h]
    \centering
    \includegraphics[width=1.0\linewidth]{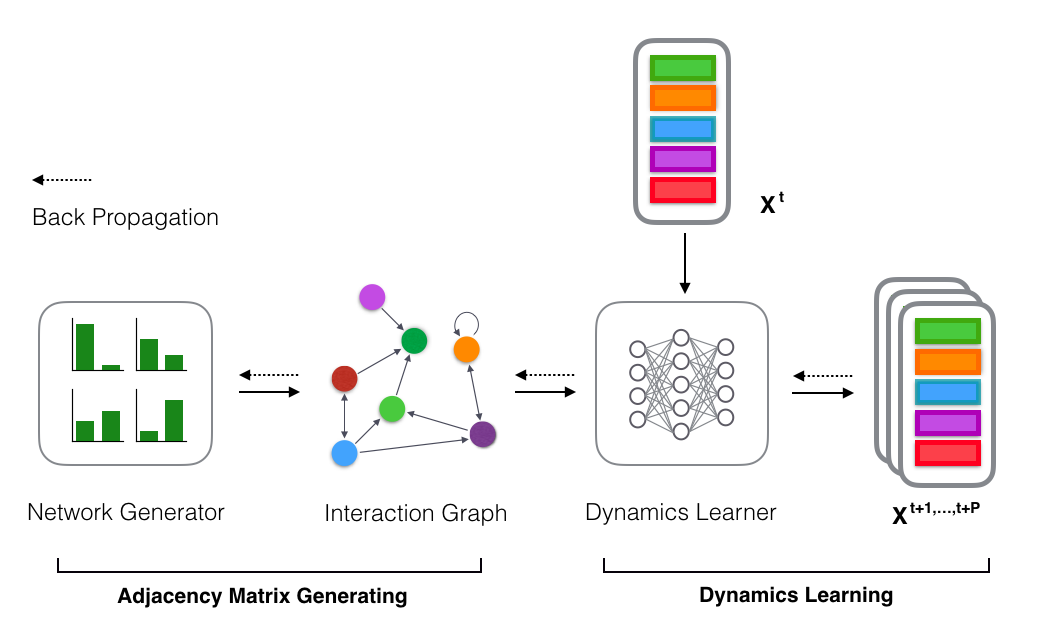}
    \caption{Basic structure of GGN. Our framework contains two main parts. First, the Adjacency Matrix is generated by the Network Generator via Gumbel softmax sampling; then the adjacency matrix and ${X}^{t}$ (node state at time $t$) are fed to Dynamics Learner to predict the node state in future $P$ time step. The back-propagation process runs back through all the computations.}
    \label{fig:structure}
\end{figure}

The general framework of our model is shown in \ref{fig:structure}. The input of the model is the feature of all nodes at time step $t$, and the output of the model is the feature of all nodes in the following $P$ steps. 
The model consists of two modules, a network generator and a dynamics learner. The job of the generator is to generate an adjacency matrix, and the learner will use the adjacency matrix generated and $X^{t}$(feature of all nodes at time $t$) to predict ${ X }^{ t+1 },...,{ X }^{ t+P }$,(feature of all nodes from time $t+1$ to $t+P$).

The Network Generator module uses the Gumbel softmax trick\cite{jang2016categorical} to generate the adjacency matrix. Details are explained in subsection 3.
The goal of the Dynamics Learner is to map the features of all nodes from time $t$ to time $t+1$ through generated adjacency matrix. Similar to NRI's design\cite{kipf_neural_2018}, our GNN comprises of 4 mapping processes between nodes and edges, which can be accomplished through MLP, CNN or RNN module. In this article, we use MLP. Details are further explained in subsection 4. To learn the complex non-linear process, we use Graph Neural Network instead of Graph Converlutional Network\cite{kipf2016semi}, since the latter does not consider the nonlinear coupling between nodes while sometimes it exists (for example, Kuramoto model).

The temporal complexity and the spatial complexity are both $O(N^2)$.

\subsection{Network Generator}

One of the difficulties for reconstructing a network from the data is the discreteness of the graph, such that the back-propagation technique, which is widely used in differential functions, cannot be applied. 

To conquer this problem, we apply Gumbel-softmax trick to reconstruct the adjacency matrix of the network directly. This technique simulates the sampling process from a discrete distribution by a continuous function such that the distributions generated from the sampling processes in real or simulation are identical. In this way, the simulated process allows for back-propagation because it is differentiable.

Network generator is a parameterized module to generate adjacency matrix. Specifically, for a network of $N$ nodes, it uses a $N \times N$ parameterized matrix to determine the $N \times N$ elements in the adjacency matrix $A$, with $\alpha_{ij}$ denoting the probability that $ A_{ij}$ takes value 1. 

Specifically, the method to generate an adjacency matrix is shown below

\begin{equation}
   A_{ij}= \frac{\exp((\log(\alpha_{ij})+\xi_{ij})/\tau)}{\exp((\log(\alpha_{ij})+\xi_{ij})/\tau)+\exp((\log(1-\alpha_{ij})+\xi'_{ij})/\tau)} \qquad, 
\end{equation}
where $\xi_{ij}$s and $\xi'_{ij}$s are i.i.d. random numbers following the gumbel distribution\cite{nadarajah2004beta}. This calculation uses a continuous function with random noise to simulate a discontinuous sampling process. And the temperature parameter $\tau$ adjusts the sharpness of the output. When $\tau\rightarrow 0$, $a_{ij}$ will take 1 with probability $\alpha_{ij}$ and 0 with probability $1-\alpha_{ij}$.

Since $\alpha_{ij}$s are all trainable parameters, they can be adjusted according to the back propagation algorithm. Thanks to the features of Gumbel-softmax trick, the gradient information can be back propagated through the whole computation graph although the process of sampling random numbers is non-differentiable.

\subsection{Dynamics Learner}

Learning with graph-structured data is a hot topic in deep learning research areas. Recently, Graph networks (GNs) \cite{battaglia2018relational} have been widely investigated and have achieved compelling performance in node classification, link prediction, etc. In general, a GN uses the graph structure $A$ and  $X^{t}$, which denotes features of all nodes at time $t$, as its input to learn the representation of each node. Specifically, the graph information used here is the adjacency matrix constructed by the generator. The whole dynamics learner can be presented as a function:

\begin{equation}
    {X}^{t}_{predict}=f({X}^{t-1},A)
\end{equation}

where $X^t$ is the state vector of all $N$ nodes at time step $t$, $A$ is the adjacency matrix constructed by the network generator.
Similar to the work \cite{kipf_neural_2018}, we realized this function through four mappings operating in succession: Node to Edge, Edge to Edge, Edge to Node and Node to Node, as shown below. Details are explained in the caption of \ref{fig:graphnetwork}.

\begin{figure}[!ht]
    \centering
    \includegraphics[width=1\linewidth]{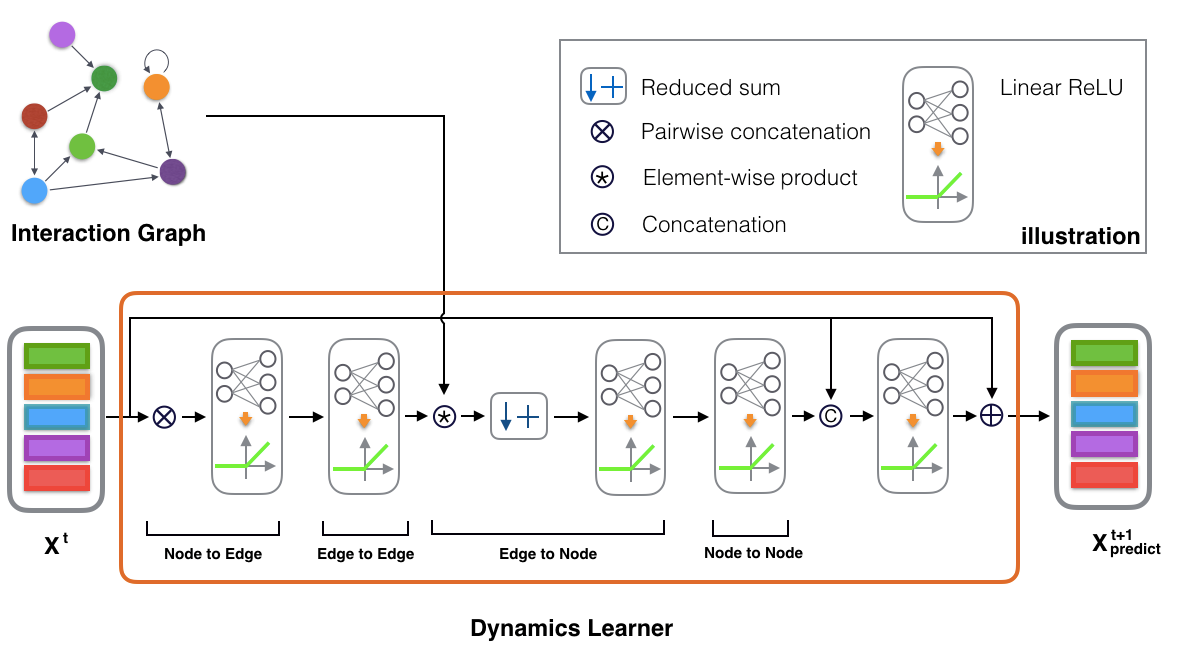}
    \caption{The Structure of the Dynamics Learner. Dynamics Learner takes the graph structure (here we use Adjacency Matrix) and node states $X$ as its input to predict node states at next time step(s). Four main parts operate in succession to accomplish the whole process: Node to Edge, Edge to Edge, Edge to Node and Node to Node. }
\label{fig:graphnetwork}
\end{figure}

\begin{equation}
H_{ e1 }^{ t-1 }=f_{ v\rightarrow e }( X^{ t-1 }\otimes({ X^{ t-1 }) }^{ T }  )
\end{equation}

\begin{equation}
H_{ e2 }^{ t-1 }=f_{ e }(H_{ e1 }^{ t-1 })
\end{equation}

\begin{equation}
H_{ v1 }^{ t }=f_{ e\rightarrow v }(A * H_{ e2 }^{ t-1 })
\end{equation}
\begin{equation}
H_{ v2 }^{ t }=f_{ v }(H_{ v1 }^{ t })
\end{equation}

Where, $H_{.}$ are hidden layers, Operation $\otimes$  is pair-wised concatenation, represented by the formula $\mathbf v \otimes \mathbf v^\top = \left\{ \langle \mathbf v_i, \mathbf v_j \rangle \right\}_{N\times N}$, resulting in a matrix where each element is a node pair. The operation is similar to the Kronecker Product except that we replace the internal multiplication with concatenation. Element-wised product * of Adjacency matrix and the result of Edge to Edge mapping sets elements 0 if there is no connection between two nodes and Reduced sum operation will aggregate edge information to the node.

Finally, we introduce skip-connection in ResNet  \cite{He2015}  to improve the gradient flow through the network, which enhances the performance of the Dynamics Learner. $X^{ t }$ denotes the nodes' states at time $t$. $f_{ output }$ is another MLP.  This process can be presented as a function

\begin{equation}
X^{ t }_{predict}=f_{ output }(\left[ X^{ t-1 },H_{ v2 }^{ t } \right] )+X^{ t-1 }
\end{equation}

Where $ \left[ .,. \right] $ denotes the concatenation operator, note that this operation, as well as the skip-connection trick are optional. We use these method only in experiments on Kuramoto. To make multi-step predictions, we feed in the output states and reiterate until we get the prediction sequence $X_{predict  }=({ X }_{ predict  } ^{ 1 } , ... ,  { X }_{predict  } ^{ T })$. Then we back propagate the loss between model prediction and the ground truth.

\subsection{Training}
Having introduced all the components, we now present the training process as algorithm below.
In the training process, we feed one step trajectory value: $X^{t}$ as its input , and their succeeding states, namely $({ X }^{ t+1 },...,{ X }^{ t+P })$ as the targets.

\begin{algorithm}[!htbp]
\SetAlgoLined
$P\leftarrow Length\:of\:Prediction\:Steps$\;
$S_{n}\leftarrow Network\:Generator\:Train\:Steps$\;
$S_{d}\leftarrow Dynamics\:Learner\:Train\:Steps$\;
$lr\leftarrow Learning\:Rate$\;

\KwIn{$X=\{{ X }^{ 0 },{ X }^{ 1 },...,{ X }^{ P }\}$}

\# Initialization:\\
Initialize: \\
Network Generator parameters $\alpha$\;
Dynamics Learner parameters $\theta$\;

\# Training:\\
 \For{each epoch}{
 $ A\leftarrow Gumbel\: Generator(\alpha)$\;
\SetSideCommentLeft{\# Train Dynamics Learner}\;
 \For{$m=1,...,S_{n}$}{
   ${ X }_{ predict  }^{ 0 }\leftarrow { X }^{ 0 }$\;
  \For{$t=1,...,P$}{
 
  ${ X }_{ predict  }^{ t }\leftarrow Dynamics\:Learner(A,{ X }_{ predict  }^{ t-1 },\theta)$\;
 
 }
 $loss\leftarrow Compute\:Loss(\{ { X }^{ 1 },...,{ X }^{ P }\} ,\{ { X }_{ predict }^{ 1 },...,{ X }_{ predict }^{ P }\} )$\;
 
 $\delta \theta \leftarrow Back Propagation$\;
 
 $ { \theta  }\leftarrow { \theta  }-lr * \delta \theta$\;
 }
 
\SetSideCommentLeft{\# Train Network Generator}\;
 \For{$n=1,...,S_{d}$}{
 
 $ A\leftarrow Gumbel\: Generator(\alpha)$\;
  ${ X }_{ predict  }^{ 0 }\leftarrow { X }^{ 0 }$\;
  \For{$t=1,...,P$}{
 
  ${ X }_{ predict  }^{ t }\leftarrow Dynamics\:Learner(A,{ X }_{ predict  }^{ t-1 }, \theta)$\;
 
 }
  $loss\leftarrow Compute\:Loss(\{ { X }^{ 1 },...,{ X }^{ P }\} ,\{ { X }_{ predict }^{ 1 },...,{ X }_{ predict }^{ P }\} )$\;
 $\delta \alpha  \leftarrow Back Propagation$\;
 
 $ { \alpha  }\leftarrow { \alpha  }-lr * \delta \alpha $\;
 }
 } 

\KwOut{$A$, $X_{predict  }=\{{ X }_{ predict  } ^{ 1 } , ... ,  { X }_{predict  } ^{ P }\}$}

 \caption{Gumbel Graph Network, GGN}
 \label{algo:GGN}
\end{algorithm}

The dynamics learner and the network generator are altering optimized in each epoch. We first optimize the dynamics learner for $S_{d}$ rounds with the network generator fixed, back propagating the loss to the dynamics learner in each round. 
Then the network generator is trained with the same loss function as the dynamics learner for $S_{n}$ rounds. In each round, the trained dynamics learner make predictions with newly generated adjacency matrix. As for loss function, when the time series to be predicted is a discrete sequence with finite countable symbols, the cross-entropy objective function is adopted otherwise the mean square errors are used. 
Note that instead of training the dynamics learner and the network generator simultaneously, we train them for different number of rounds per epoch. The main reason behind this training process is the observation that the dynamics and network structures evolve in different paces in real systems. Networks usually change slower than the dynamics. Small changes on network structure may lead to dramatic changes on node dynamics. Therefore, by alternating the training processes in different rounds per epoch, we can adjust both learning processes to appropriate paces. 

In practice, $S_{d}$ and $S_{n}$ vary case by case. Although we chose them mainly through hyper-parameter tuning, there is a general observation that the more complex the dynamics is, the larger $S_{d}$ it requires. For example, for Boolean Network model mentioned below, which exhibiting binary dynamics, the $S_{d}$ is 10; while for Kuramoto model, which is highly nonlinear, the $S_{d}$ needs to be around 30 to achieve a good result.

\section{Experiments}
\label{section-exp}
\subsection{An Example}

At first, we will show how GGN works and in what accuracy, we use a 10-body mass-spring interaction system as an example. Suppose in a two-dimensional plane, there are 10 masses linked each other by springs, and the connection density is $0.2$. The masses can move according to the spring dynamics if the initial positions and velocities are given. And we will use the data of the position and velocity of each particle generated by the simulation to reconstruct their connections and predict their future positions.

In this experiment, we set $S_n=5$ and $S_d=50$, and use 5k training samples,1k validation samples and 1k test samples with each sample containing 10-steps trajectory of each mass. In each sample, one initial condition is adopted. The result shows that GGN can reconstruct the adjacency matrix with 90\% accuracy, and can predict the next step positions with a fairly small error $2.97e-5$. Figure \ref{fig:structure} visualizes the trajectories of simulation model and predictions.

%\begin{table}[!h]
%\caption {Results with spring-connected particles.}
%\centering
%\begin{tabular}{|c|c|c|c|}
%\hline
%ACC(\%)&TPR&FPR&MSE      \\\hline

%90&0.727&0.044&2.97e-5  \\\hline

%\end{tabular}
%\label{tab:results-spring}
%\end{table}

\begin{figure}[!h]
    \centering
    \includegraphics[width=1.0\linewidth]{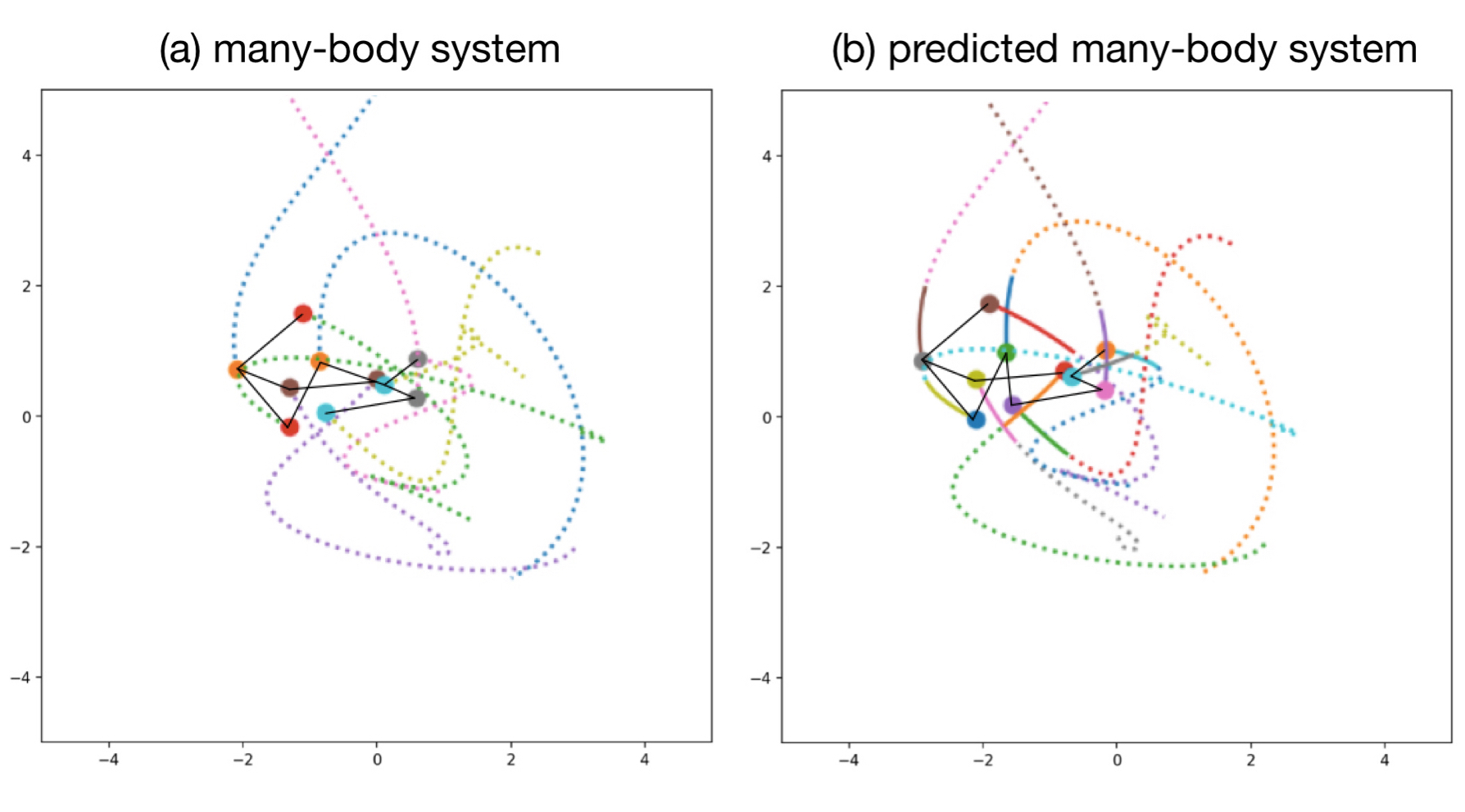}
    \caption{Real and predicted states of 10-body mass spring system.The two schematics show a true and predicted state of the mass spring system. Figure (a) shows the trajectories of the interactions of masses and the connections between them, and Figure (b) shows the future state (solid line) and connection relationships that we predict using the data (dotted line trajectory)}
    \label{fig:structure}
\end{figure}

It can be seen that GGN model can reconstruct the many-body problem in two-dimensional plane with high accuracy, and it can predict the node state in the future time accurately.

\subsection{Experiments on Simulated Models}
To systematically test the power of GGN, we experimented it on three types of  simulated models: Boolean Network\cite{kauffman1969metabolic}, Kuramoto \cite{kuramoto1975self}, and Coupled Map Lattice \cite{kaneko1992overview, kaneko1989pattern}, which exhibit binary, continuous, and discrete trajectories, respectively. Here we attempt to train our model to learn the dynamics and reconstruct the interactions between particles, or the adjacency matrices, under all three circumstances.

Furthermore, we test the performance of GGN on different parameters with three main experiments: one concerns different net size and different level of chaos (subsection 3.1); one features different type of network topology (subsection 3.2), and one studies the relationship between data size and accuracy (subsection 3.3). Our full code implementations are as shown on Github [https://github.com/bnusss/GGN].

\begin{figure}[h!]
\centering
\includegraphics[scale=0.1]{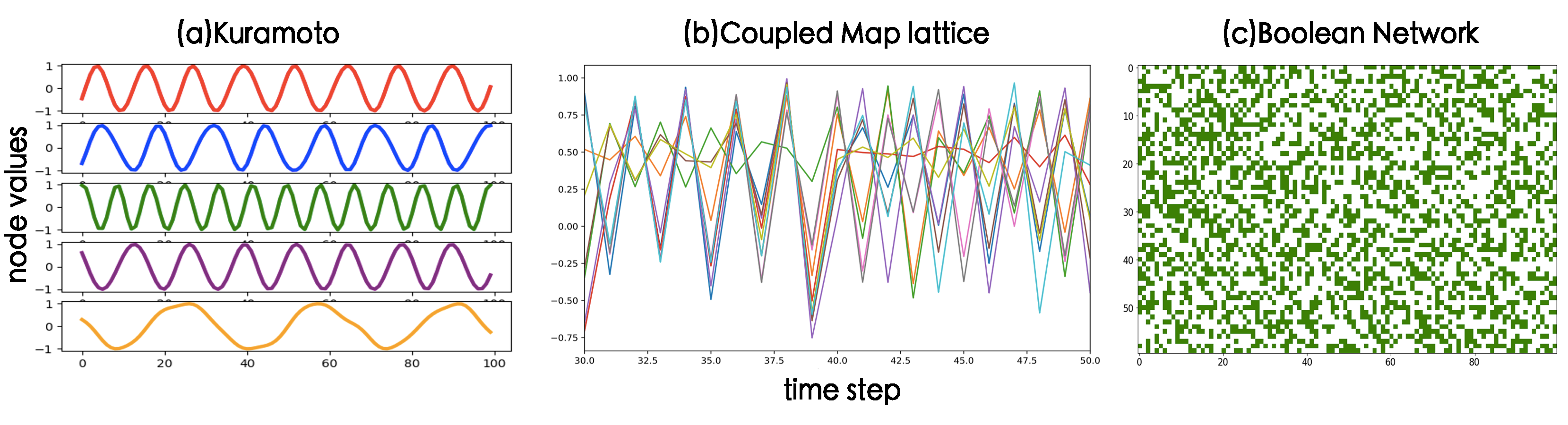}
\caption{Time evolution of node values in three types of network simulations models. Kuramoto (a), Coupled Map Lattice (b) and Boolean Network (c).}
\label{fig:threekinds}
\end{figure}

\subsubsection{Boolean Network}
% BN in this experiment...
%% 模型概述
Boolean Network is a widely studied model that is often used to model gene regulatory networks. In Boolean Network system, every variable has a possible value of 0 or 1 and a Boolean function is assigned to the node. The function takes the states of its neighbors as inputs and returns a binary value determining the state of the current node.

In simulation. We set the structure of the network as a directed graph with the degree of each node as $K$, and different $K$ determines whether the network will evolve chaotically or non-chaotically. All nodes follow the same randomly generated table of dynamical rules. The training data we generated contains 5k pairs of state transition sequences. Meanwhile, we simulated 1k validation set and 1k test set.

\subsubsection{Kuramoto Model}
The Kuramoto model \cite{kuramoto1975self} is a nonlinear system of phase-coupled oscillators, and it is often used to describe synchronization. Specifically, we study the system 
\begin{equation}
    \frac{d \phi_i}{dt} = \omega_i + k\sum_{j \neq i} A_{ij} \sin(\phi_i - \phi_j)
\end{equation}

Where $\omega_i$ are intrinsic frequencies sampled from a given distribution $g(\omega )$, and here we use a uniform distribution on $[1, 10)$; $k$ is the coupling strength; $A_{ij}\in \left\{ 0,1 \right\} $ are the elements of $N \times N$ adjacency matrix, and for undirected random networks we study, $A_{ij}= A_{ji}$. The Kuramoto network have two types of dynamics, synchronized and non-synchronized. According to studies by Restrepo et. al \cite{restrepo2005onset}, the transition from coherence to incoherence can be captured by a critical coupling strength $k_{c}= k_{0}/{\lambda}$, where $k_{0}=2/{{\pi}g(5.5)}$ in our case, and $\lambda$ is the largest eigenvalue of the adjacent matrix. The network synchronizes if $k>k_{c}$, and otherwise fails to synchronize. We simulate and study both coherent and incoherent cases. 

For simulation, we solve the 1D differential equation with 4th-order Runge-Kutta method with a step size of $0.01$. Our training sets include 5k samplings, validation set 1k, and test set 1k, each sampling covers $d\phi_i / dt$ and $\sin(\phi_i)$ in 10 time-steps.

\subsubsection{Coupled Map Lattice}
% CML setting in this experiment...
%% 模型概述
Coupled map lattices represent a dynamical model with discrete time and continuous state variables\cite{kaneko1992overview, kaneko1989pattern},it is widely used to study the chaotic dynamics of spatially extended systems. The model is originally defined on a chain with a periodic boundary condition but can be easily extended to any type of topology:
\begin{equation}
    x_{t+1}(i)=(1-s)f(x_{t}(i))+\frac{s}{\text{deg}(i)}\sum_{j \in \text{neighbor}(i)} f(x_{t}(j)),
\end{equation}
where $s$ is the coupling constant and $\text{deg}(i)$ is the degree of node $i$. We choose the following logistic map function:
\begin{equation}
    f(x) = \lambda x(1-x).
\end{equation}

%% Experiment setup
We simulated $N \in \{10, 30\}$ coupled map lattices with initial states $x_0(i)$ sampling uniformly from $[0, 1]$ for random regular graphs. Notice that when setting coupling constant $s=0$, the system reduces to $N$ independent logistic map. The training sets also include 5k samplings, 1k validation set, and 1k test set, each sampling covers $x_i$ in 10 time-steps.

\section{Results}

In each experiment listed below, we set the hyper-parameters $S_{n}$ and $S_{d}$ of the Boolean Network model to $20$ and $10$, respectively, while in Coupled Map Lattice model and Kuramoto model they are $5$ and $30$. In Coupled Map Lattice model and Kuramoto model, the prediction steps $P$ is $9$, which means that the current state is used to predict the node state of the next $9$ time steps, while in the Boolean Network, it is set to $1$. In all the experiments, we've set the hidden size in all the MLP networks of the dynamics learner module of the GGN model to $256$. All the presented results are the mean value over five times of repeated experiments. The horizontal lines: ``-'' in the table indicates that the amount of data exceeds the model processing limitation, the model becomes so unstable that outputs may present as ``nan'' during training.

We compare our model with following baseline algorithms:

\begin{itemize}
\item \textbf{LSTM}(Long Short-Term Memory Network) is a well-known recurrent neural network and has been shown to be very suitable for sequence prediction problems. To do network reconstruction with LSTM,  previous work \cite{kipf_neural_2018} used thresholded correlation matrix to represent the adjacency matrix. {But according to our experiments, this method would only yield all-zero or all-one matrices, therefore cannot serve as a satisfactory way of deriving adjacency matrices. Hence, we use LSTM only for node state prediction.}However, this method cannot obtain meaningful results as in \cite{kipf_neural_2018} because different network generating methods are used. Therefore, we ignore the network reconstruction accuracy while only the state prediction is reported for LSTM. %while Kipf et.al have different adjacency matrices for each sampling, we study the network of only one adjacency matrix throughout one train-validate-test process, so if we are to calculate the threshold value basing on validation set as well, then it means we are using information from test set, which renders testing totally pointless. Hence, in this article, we are unable to adopt this method, and only use LSTM only for node state prediction.%

\item \textbf{NRI}(Neural Relational Inference Model) is able to reconstruct the underlying network structure and predict the node state in future time steps simultaneously by observing the node state sequence. We compare our model against it in both tasks. Here we use settings similar to that in Kipf's original paper\cite{kipf_neural_2018}: all our experiments use MLP decoders, and with the Kuramoto model, we use CNN encoder and other models the MLP encoder.
\end{itemize}

We use the following indicators to evaluate the results of the experiments:

\begin{itemize}

\item \textbf{TPR}(true positive rate) measures the proportion of positive instances that are correctly identified. We consider 1 in the adjacency matrix as a positive element, whereas 0 as a negative one.

\item \textbf{FPR}(false positive rate) computes the proportion of negative instances that are incorrectly identified in the adjacency matrix generated.

\item \textbf{MSE}(mean square error) measures the average of the squares of the errors, that is the average squared difference between the estimated values and data. The MSE we showed below is the average mean square error of next $P$ time steps.

\item \textbf{ACC(net)} is the proportion of correctly identified elements of the Adjacency Matrix.

\item \textbf{ACC(dyn)}, In our experiment on the Boolean Network, we use indices ACC(dyn) to measure the proportion of nodes whose states are predicted accurately in the next time step.
\end{itemize}

\label{section-results}
\subsection{Experiments with Different Dynamics}

\begin{table}[h]
\caption{Results with Boolean Network.}
% \centering
\begin{tabular}{|c|c|c|c|c|c|c|c|c|c|c|}
\hline
Node Num &State  &LSTM& \multicolumn{4}{|c|}{NRI}& \multicolumn{4}{|c|}{GGN}\\ \cline{3-11} 
&&ACC(dyn)&ACC(net)(\%)&TPR&FPR&ACC(dyn)  & ACC(net)(\%)&TPR&FPR&ACC(dyn)  \\\hline
10 & non-chaotic   & \textbf{0.841} & 56.8&0.422&0.395&0.820 & \textbf{99.1} & 1 & 0.008&0.694  \\\hline
10 & chaotic &\textbf{0.789} & 48.1 &0.458    & 0.465 & 0.528  & \textbf{99.4} & 0.983 & 0&0.693  \\\hline
30 & non-chaotic   &0.912 &40.9&0.590& 0.591&0.798 & \textbf{92.6} & 0.476 & 0.036&\textbf{0.948}  \\\hline
30 & chaotic & \textbf{0.765}&46.0&0.549&0.547&0.721 & \textbf{90.0}& 0.601 & 0.034&0.699  \\\hline
100 &non-chaotic   & 0.933&-&-&-&-& \textbf{84.0}  & 0.505 & 0.153&\textbf{0.982}  \\\hline
100 &chaotic & \textbf{0.796}&-&-&-&-& \textbf{95.7} & 0.25  & 0.013&0.7483  \\
\hline
\end{tabular}
\label{tab:results-bn}
\end{table}

In our experiments, we set the network topology of a Boolean Network as a directed graph, with the indegree of each node being $k$, and $k$ determines whether the system is chaotic or not. For all systems, we set $k=2$ for non-chaotic cases; and for systems with $10, 30, 100$ nodes, $k$ is set to $7, 5$ and $4$, respectively, to obtain chaotic dynamics. As shown in Table {\ref{tab:results-bn}}, the GGN model recovers the ground-truth interaction graph with an accuracy significantly higher than competing method, and the recovery rate in non-chaotic regimes is better than those in chaotic regime.

\begin{table}[h]
\caption{Results with CML model.}
\centering
\begin{tabular}{|c|c|c|c|c|c|c|c|c|c|c|}
\hline
Node Num & State&LSTM & \multicolumn{4}{|c|}{NRI}& \multicolumn{4}{|c|}{GGN}\\ \cline{3-11}

&&MSE& ACC(\%) & TPR & FPR&MSE& ACC(\%) & TPR & FPR&MSE   \\ \hline
10       & non-chaotic &1.92e-2&53.1&0.446&0.588 &1.69e-4& \textbf{100}          & 1   & 0 & \textbf{5.63e-6}   \\ \hline
10       & chaotic    &2.54e-2&54.7&0.459&0.605&4.04e-4& \textbf{99.3}          & 1   & 0.013  & \textbf{3.24e-5}  \\ \hline
30       & non-chaotic&4.11e-2 &-&-&-& -&\textbf{100}         & 1   & 0 & \textbf{3.29e-6} \\ \hline
30       & chaotic    &5.03e-2 &-&-&-& -&\textbf{99.9}         & 1   & 0.0017 & \textbf{3.41e-6} \\ \hline
\end{tabular}
\label{tab:results-cml}
\end{table}

Here we presented our results obtained on coupled map lattice model in Table \ref{tab:results-cml}. In our experiments, the network topology is random 4-regular graph, and we set coupling constant $s=0.2$ fixed. Because $r \approx  3.56995$ is the onset of chaos in the logistic map, we chose $r=3.5$ and $r=3.6$ to represent non-chaotic and chaotic dynamics respectively. For a random 4-regular graph with 10 nodes, our GGN model has obtained approximately 100\% accuracy in the task of network reconstruction. For a system with 30 nodes, it is still able to achieve a high accuracy and the performance obtained on non-chaotic dynamics is better than that on chaotic dynamics.

\begin{table}[h]
\caption {Results with Kuramoto model.}
\centering
\begin{tabular}{|c|c|c|c|c|c|c|c|c|c|c|c|c|c|}
\hline
Node Num &State &LSTM & \multicolumn{4}{|c|}{NRI}& \multicolumn{4}{|c|}{GGN} \\ \cline{3-11}
&&MSE&ACC(\%)&TPR&FPR&MSE  & ACC(\%)&TPR&FPR&MSE      \\\hline

10 & coherent   &2.67e-2&51.8&0.543&0.505 &9.63e-2& \textbf{99.8} & 0.994 & 0.004&\textbf{1.12e-3}  \\\hline
10 & incoherent &2.85e-2&51.7&0.543    & 0.508    &1.11e-1& \textbf{99.8} & 0.994 & 0.004&\textbf{8.36e-4}  \\\hline
30 & coherent   &3.12e-2&-&-&-&-& \textbf{89.8} & 0.920 & 0.169&\textbf{3.96e-4}  \\\hline
30 & incoherent &3.35e-2&-&-&-&-& \textbf{81.0} & 0.700 & 0.124& \textbf{1.90e-4}  \\\hline
%100 &coherent   &-&-&          & 0.84  & 0.505 & 0.153 &-&-& - \\\hline
%100 &incoherent &-&-&          & 0.957 & 0.25  & 0.013 &-&-& - \\\hline
\end{tabular}
\label{tab:results-kuramoto}
\end{table}

In the experiment concerning Kuramoto Model, we used Erdos-Renyi random graph with a connection possibility of $0.5$. As the onset of synchronization is at $k=k_{c}$ (in our cases, $k_{c}=1.20$ for 10 nodes system and $k_{c}=0.41$ for 30 nodes system), we chose $k=1.1k_{c}$ and $k=0.9k_{c}$ to represent coherent and incoherent dynamics respectively. Here we used data of only two dimensions (speed and amplitude), as opposed to four dimensions in NRI's original setting (speed, phase, amplitude and intrinsic frequency), so the performance of NRI model here is lower than that presented in Kipf's original paper \cite{kipf_neural_2018}. Similar to BN and CML, our GGN model attains better accuracy in coherent cases, which are more regular than incoherent ones.

To sum up, the above experiments clearly demonstrates the compelling competence and generality of our GGN model. As shown in the tables, GGN achieves high accuracy on all three network simulations, and its performance remains good when the number of nodes increases and the system transform from non-chaotic to chaotic. Although we note that our model achieves relatively lower accuracy in chaotic cases, and it is not perfectly stable under the current data size (in CML and Kuramoto experiments, there is a 1/5 chance that performance would decrease), the overall results are satisfactory.

\subsection{Reconstruction Accuracy with Network Structure}

In this section, we used 100-node Boolean Network with the Voter dynamics \cite{li2015reconstructing} (for a node with degree $k$, and $m$ neighbours in state 1, in the next time step it has a probability of $m/k$ to be in state 1, and a probability of $(k-m)/k$ to be in state 0) to study how network structure affects the network reconstruction performance of our GGN model. Specifically, we studied WS networks and BA networks, and examined how the reconstruction accuracy would change under different network parameters. We also experimented with two different data sizes: 500 and 5000 pairs of state transition sequences, to see how network structure would affect the needed amount of data.

\floatsetup[figure]{style=plain,subcapbesideposition=top}

    \begin{figure}[htb]\centering
\sidesubfloat[]{\includegraphics[width=0.48\textwidth]{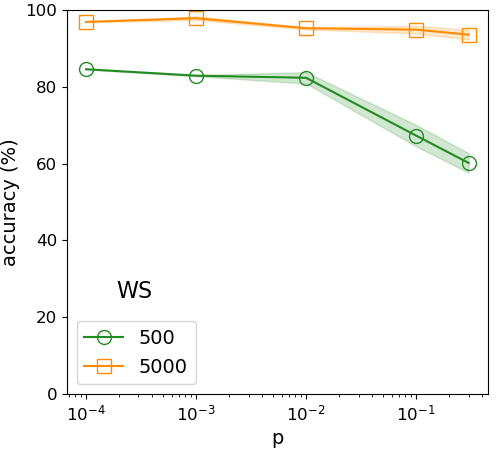}\label{fig:a}}
\sidesubfloat[]{\includegraphics[width=0.48\textwidth]{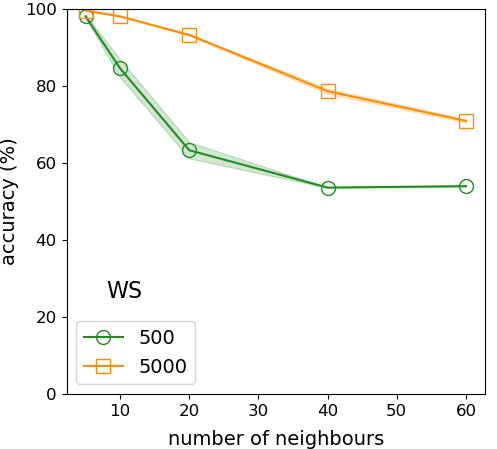}\label{fig:b}}

\sidesubfloat[]{\includegraphics[width=0.48\textwidth]{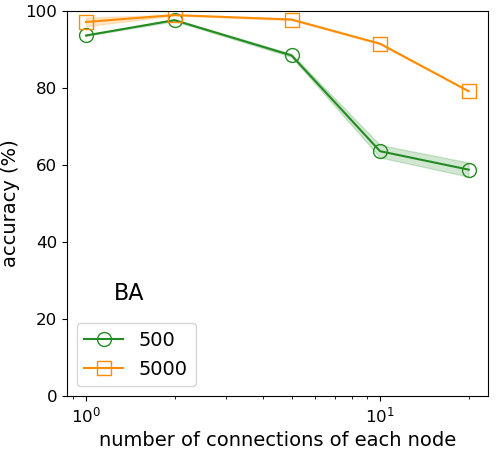}\label{fig:c}}
%\hfil
\caption{Accuracy of reconstruction with different network structures. Accuracy of reconstruction with \textbf{a}: WS networks under different re-connection possibility $p$ (while neighbours=20); \textbf{b}: WS networks under different number of neighbours (while $p=0.3$); \textbf{c}: BA networks under different number of connections of each node. We experimented with two different data sizes: 500 and 5000 pairs of state transition sequences, represented in each plot by green and orange line, respectively.}
    \label{fig:myfigure}
    \end{figure}

 In the first experiment, we studied WS networks of different re-connection possibility $p$. We note that the reconstruction accuracy declines slowly between $p=$ $[10^{-4}, 10^{-2}]$, but drops sharply when $p$ is larger than $10^{-2}$. As the average distance of the network drops quickly before $10^{-2}$, but our reconstruction accuracy remains roughly the same, it seems that the reconstruction accuracy is insensible to it. On the other hand, the Clustering Coefficient of the network drops quickly when $p$ is larger than $10^{-2}$, while declining slowly when $p$ is smaller \cite{watts1998collective}, which correspond with our curves of accuracy. Therefore, we may conclude that the reconstruction accuracy is directly affected by Clustering Coefficient of the network.  
 However, as the data size increases, the performance is significantly augmented under all different values of $p$, and the slope is greatly reduced. So increasing the data size can effectively solve the problem brought by increasing re-connection possibility.

 In the second experiment, we studied WS networks of different number of neighbours. Here the situation is much simpler: as number of neighbours increases, the complexity of network also goes up, and it in turn makes learning more difficult. So the need for data is increasing along with the number of neighbours.

In the third experiment, we studied BA networks of different number of connections of each node. The result is similar to the second experiment, but here, increasing the data size receives a smaller response than in the WS networks. That is probably because in BA networks, a few nodes can greatly affect the dynamics of the whole network, which makes the complexity even higher, therefore the need for data would be greater for BA networks.

\subsection{Reconstruction Accuracy with Data Size}
\begin{figure}[!htbp]
    \centering
    \includegraphics[width=0.8\linewidth]{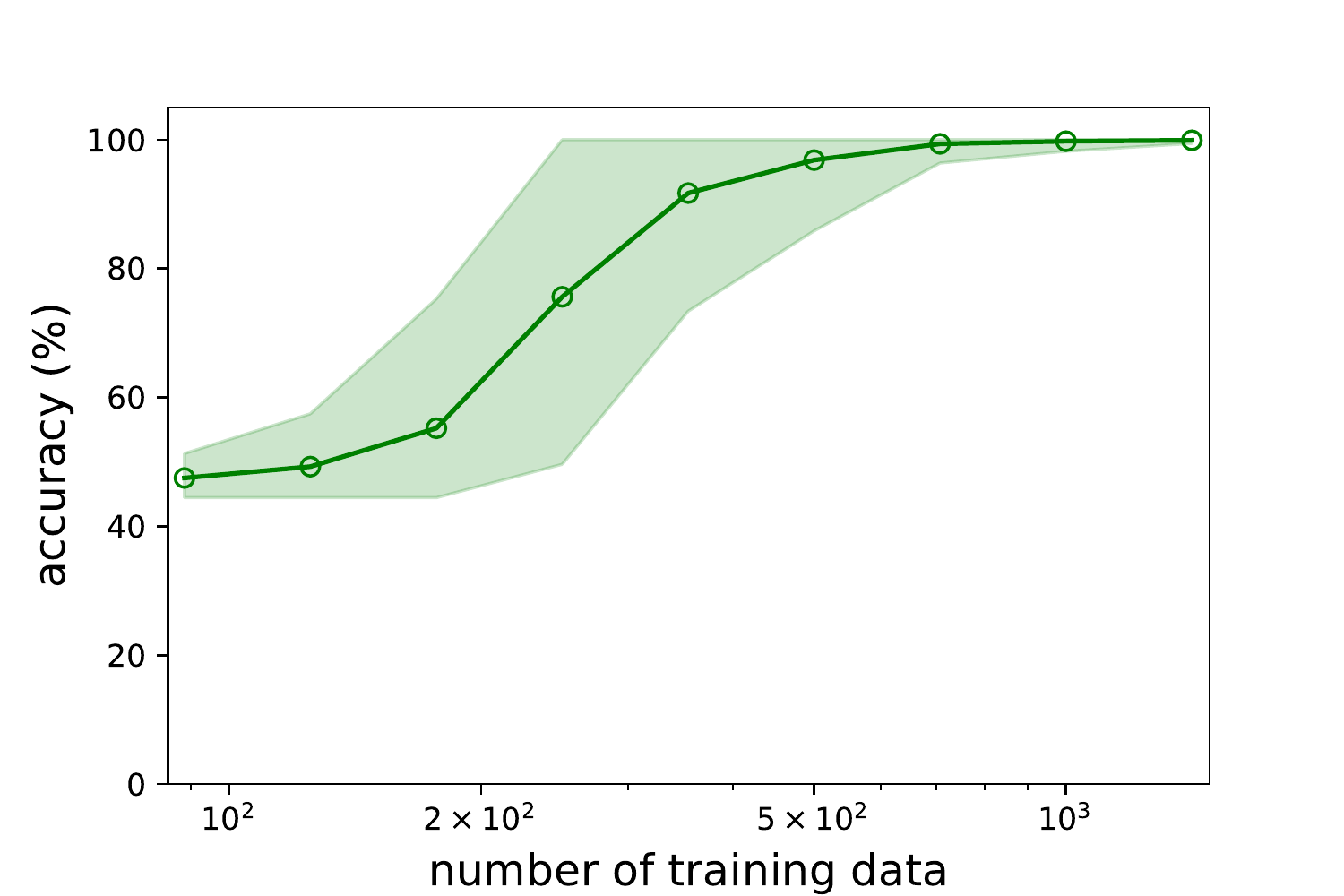}
    \caption{Accuracy of reconstruction versus the size of training data. The amount of training data is ranging from about $10^2$ to $10^3$ with each sampling evolving for $T=100$ time steps. The results are obtained on CML model with 10 nodes and the network topology is random 4-regular graphs. We set $s=0.2$ and $r=3.6$ to generate chaotic dynamics.}
    \label{fig:acc-vs-num-data}
\end{figure}

In this section we study the dependency between the amount of data and the accuracy of reconstruction. We performed our experiments on CML model with chaotic dynamics. As illustrated in Figure \ref{fig:acc-vs-num-data}, the accuracy of reconstruction significantly improves when feeding more data to the model. We also noticed that an insufficient amount of data can lead to a high deviation, which means that our method can either produce results with high accuracy or fail.

\section{Conclusion}
\label{section-conclusion}

%%%第一版，主实验放在第二段，强调不需要先验知识
In this work we introduced GGN, a model-free, purely data-driven method that can simultaneously reconstruct network topology and perform dynamic prediction from time series data of node state. Without any prior knowledge of the network structure, it is able to complete both tasks with high accuracy. 

In a series of experiments, we demonstrated that GGN is able to be applied to a variety of dynamical systems, including continuous, discrete, and even binary ones. And we found that in most cases, GGN can reconstruct the network better from non-chaotic data. In order to further explore GGN's properties and to better know its upper limit, we conducted experiments under different network topology and different data volumes. The results show that the network reconstruction ability of our model is strongly correlated with the complexity of dynamics and the Clustering Coefficient of the network. It is also demonstrated that increasing the data size can significantly improve GGN's net reconstruction performance, while a small data size can result in large deviation and unstable performance.

However, we are well aware that the current work has some limitations. It now focuses only on static graph and Markovian dynamics. Besides, as the limitation of our computation power, the maximum network size we can process is limited up to 100 nodes. Several possible approaches may help us to conquer these problems. First, if we can parameterize the network generator in a dynamical way, that is, to allow the generator parameters change along time, we can break through the limitations of static graphs. Second, if we replace the MLP network with RNN in the framework, learning of non-Markovian dynamics is possible. Third, to improve the scalability of our framework, node by node reconstruction of network can be adopted to save the space complexity. Another good way to improve the size limitation is to use graph convolution network (GCN) to model dynamics learner. GCN has been proved to be very useful in a large variety of field although it can not simulate some complex nonlinear process quite well from the experimental.%
 In future works, we will further enhance the capacity of our model so as to break through these limitations.

\section{Declarations}
\subsection{Availability of data and material}
The datasets generated and/or analysed during the current study are available in https://github.com/bnusss/GGN.

\subsection{Competing interests}
The authors declare that they have no competing interests.

\subsection{Funding}
The research is supported by the National Natural Science Foundation of China (NSFC) under the grant numbers 61673070.

\subsection{Authors' contributions}
JZ conceived and designed the study. ZZ, YZ, SW, JL, RT and RX performed the experiments. ZZ, YZ, SW, JL, RT and RX wrote the paper. JZ reviewed and edited the manuscript. All authors read and approved the manuscript.

\subsection{Acknowledgements}
% \subsection{Authors' information (optional)}
We thank professor Wenxu Wang and Qinghua Chen from School of Systems Science, Beijing Normal University for discussion and their generous help.

\bibliographystyle{bmc-mathphys} 
\bibliography{references}

\end{document}